\documentclass[conference]{IEEEtran}
\IEEEoverridecommandlockouts
\usepackage{cite}
\usepackage{amsmath,amssymb,amsfonts}
\usepackage{graphicx}
\usepackage{textcomp}
\usepackage{xcolor}
\usepackage{booktabs}   %
\usepackage{array}

\begin{document}

\title{Beyond Accuracy: What Are Speech Emotion Recognition Models Learning?}

\author{\IEEEauthorblockN{Prince Baah-Mensah}
\IEEEauthorblockA{
\textit{Ashesi University}\\
prince.baah@ashesi.edu.gh}
\and
\IEEEauthorblockN{Dennis Asamoah Owusu}
\IEEEauthorblockA{
\textit{Ashesi University}\\
dowusu@ashesi.edu.gh}
}
\maketitle

\begin{abstract}
Speech Emotion Recognition (SER) is important in human-machine interactions, enabling machines to detect and classify human emotions from solely human speech.
Speech emotion recognition (SER) benchmarks routinely report accuracies above 90 percent, yet these models fail when deployed or tested on diverse speakers. The papers  overlook a fundamental issue: models may be learning speaker identity and tone rather than generalized emotional cues. We introduce a two-condition
audit protocol that measures, for any published SER model, how much of its
reported performance survives when speakers are held out. Each model is
evaluated twice under identical architecture, features, and training recipe: a  control condition that reproduces the common random-split practice, and a strict speaker-independent condition with disjoint speakers. We apply the protocol to seven model configurations
from five published papers, spanning convolutional, recurrent, transformer,
vision-transformer, and self-supervised wav2vec2 architectures, on RAVDESS,
EMO-DB, and SAVEE, with a fourth corpus (EARS) as an external check. Across
21 model-dataset pairs, published headline numbers reproduce close to the
control but drop by 6.9 to 63.1 percentage points under strict speaker
independence. Our results quantify speaker
memorization as a dominant, previously unmeasured component of published SER
performance, and we recommend two-condition reporting as a community standard.\end{abstract}


\section{Introduction}\label{sec:intro}
Speech emotion recognition (SER) has become a benchmark-driven field.
Successive architectures report ever-higher accuracies on a small set of 
corpora, and these numbers now serve as evidence of progress
\cite{schuller2018}, \cite{scheidwasser2022}. Comparisons across papers,
however, assume that the reported numbers measure the same quantity: the
ability to recognize emotion from speech.

That assumption hides a second variable. In training and evaluation, emotion and speaker
identity are mixed: each actor appears in both the training and test sets. Any evaluation in which test speakers
also appear in training therefore admits a shortcut, recognizing the speaker
rather than the emotion. Common practice enables this shortcut in three ways:
random train/test splits that ignore speaker boundaries, speaker normalization
computed over the full corpus including test utterances, and data augmentation
applied before splitting, which duplicates the same speakers across both sides
of the boundary \cite{ahmed2023}, \cite{kumar2022}, \cite{pepino2021}.

The consequence is visible but unquantified. Papers that adopt speaker-disjoint
protocols report markedly lower numbers than those that do not; Li et al., who
evaluate under a session-disjoint protocol, report weighted accuracy of 62.16
percent on IEMOCAP, where leaked-split studies on comparable corpora exceed 90
percent \cite{li2021}. Domain-adaptation work has long treated speaker as a
domain and shown that speaker-disjoint evaluation changes model rankings
\cite{xi2019}. Yet no study has measured, for the same architecture and
training recipe, how much of a published number survives speaker hold-out.
Without that measurement, the field cannot distinguish emotion learning from
speaker memorization.

We provide that measurement. We define a two-condition audit protocol. In the
control condition we reproduce each paper's evaluation practice: random stratified splits with speakers shared across
train, validation, and test, together with the paper's own augmentation. In the
strict speaker-independent (SI) condition we hold out all test speakers, choose
validation speakers disjoint from both, and compute all normalization
statistics from training data only. Every configuration is trained and
evaluated under both conditions with identical hyperparameters, and we report accuracy and unweighted average recall (UAR) as mean and standard
deviation over three control seeds and all SI folds. We audit seven configurations from five papers: wav2vec2 with a dense head
\cite{pepino2021}, \cite{baevski2020}; the EmoFormer CNN-transformer
\cite{hasan2024}; three configurations from Ahmed et al., comprising a baseline
1D-CNN with fully connected layers (Model A) and its variants augmented with
LSTM (Model B) and GRU (Model C) recurrent heads \cite{ahmed2023}; the
BLSTM-DSA of Li et al. \cite{li2021}; and the mel-spectrogram vision
transformer of Kumar et al. \cite{kumar2022}. Datasets are RAVDESS
\cite{livingstone2018}, EMO-DB \cite{burkhardt2005}, and SAVEE
\cite{haq2014}, with EARS \cite{ears2023} as a fourth-corpus check for the one
model published on it.

The audit yields three findings. First, published headline numbers are
reproducible, but only under the leaked condition: our control runs recover the
claimed accuracies within a few points for every paper whose protocol we could
replicate. Second, the same models lose 6.9 to 63.1 percentage points of
accuracy when speakers are held out. Third, the collapse tracks how much the model learned under leakage: from-scratch models with high control accuracy suffer severe drops under strict SI, whereas low-gap from-scratch models simply underfit both conditions.

Our contributions are methodological and quantitative. (1) A two-condition
audit protocol that isolates speaker memorization from emotion learning for any
published SER model. (2) A quantitative audit of seven configurations from five
papers across three corpora and large-scale corpus, with per-fold and
per-seed statistics. The remainder
of the paper is organized as follows. Section II reviews benchmark practice and
the domain-shift view of speaker variability. Section III defines the protocol.
Section IV reports the audit. Section V discusses sources of memorization, what
survives strict evaluation, and limitations. Section VI concludes.

\section{Related Work}\label{sec:related}
Most speech emotion recognition (SER) research relies on datasets like RAVDESS \cite{livingstone2018}, EMO-DB \cite{burkhardt2005}, and SAVEE \cite{haq2014}. In these datasets, the same actors record every emotion. This creates a hidden shortcut: a model can simply memorize the actor's unique voice rather than learning the actual emotion. 

Many published papers accidentally allow this shortcut through their testing methods. For example, Ahmed et al.\ \cite{ahmed2023} split their data randomly 80/20, Kumar et al.\ \cite{kumar2022} use a random 64/20/16 split, and the EmoFormer paper \cite{hasan2024} uses a random 70/30 split. In all of these cases, the same speakers appear in both the training and testing sets, which leaks speaker identity. Even when papers try to separate speakers, they sometimes leak information in other ways. For instance, Pepino et al.\ \cite{pepino2021} use a speaker-disjoint split but also report a speaker-normalized version that calculates audio statistics using the test speakers.

Some researchers use domain adaptation techniques to remove speaker traits from the audio, or rely on self-supervised pre-training \cite{baevski2020} to learn speaker-invariant features. Others use strict testing rules, like the leave-one-session-out method used by Li et al.\ \cite{li2021}, to ensure test speakers are never seen during training. However, these tests are almost never compared side-by-side with a leaked test using the exact same model. While surveys \cite{schuller2018} and multi-corpus benchmarks \cite{scheidwasser2022} standardize features and models, they do not standardize how speakers are split. 

Because no one runs both the leaked and honest tests on the same models, the field does not know how much of a paper's high accuracy is actually just speaker recognition. Our work fills this gap by running both conditions on the exact same architectures to measure the cost of speaker leakage.

\section{Audit Protocol}\label{sec:protocol}
\subsection{Two Evaluation Conditions}
Every audited configuration is evaluated under two conditions that differ only
in how speakers are partitioned; architecture, features, augmentation,
optimizer, and early-stopping rule are identical across conditions.

\emph{Control (leaked).} We reproduce the evaluation practice common to the
audited papers: a random stratified split into train, validation, and test in
which the same speakers appear on all three sides, combined with each paper's
own augmentation pipeline. Three seeds per dataset (five for the wav2vec2
configuration, following the audited paper's seed list) instantiate the split,
and we report mean and standard deviation across seeds.

\emph{Strict speaker-independent (SI).} Test speakers are disjoint from
training speakers: grouped six-fold cross-validation on RAVDESS (four test
speakers per fold), leave-one-speaker-out on EMO-DB and SAVEE, and grouped
five-fold on EARS. In every fold the validation speaker set is disjoint from
both train and test. All dataset-level normalization statistics (feature-wise
means and standard deviations) are estimated on training data only and applied
unchanged to validation and test; per-clip operations such as spectrogram
min-max scaling are leakage-free by construction. One run per fold; we report
mean and standard deviation across folds.

Two metrics summarize the audit. The speaker-memorization gap is the leaked accuracy minus the SI accuracy, and retention is the ratio of SI accuracy to leaked accuracy. However, the size of the accuracy drop does not tell the whole story. A poorly performing model might show a very small drop simply because it failed to learn the training data in the first place. Therefore, to truly understand a model's behavior, we must evaluate the size of the drop alongside its baseline score in the leaked condition.\subsection{Datasets}

Table~\ref{tab:datasets} lists the corpora. RAVDESS, EMO-DB, and SAVEE are
corpora in which a small number of speakers express every emotion, the
setting in which speaker and emotion are most strongly confounded; we retain
each corpus's full emotion set (8, 7, and 7 classes) so that chance levels are
uniform across models. EARS, an anechoic expressive-speech corpus with 107
speakers, serves as a fourth check for the one audited model
published on it. Speaker counts range from 4 (SAVEE) to 107 (EARS), which
makes the SI condition progressively harder and exposes how each configuration
behaves as the speaker diversity available during training shrinks. Two of the audited papers evaluate EMO-DB on a reduced label set: Kumar et al.\
and Li et al.\ report results on only the four extreme emotions (anger,
happiness, sadness, neutral). Because chance level and task difficulty change
with the number of classes, a cross-model comparison requires an identical
label set. We therefore re-ran these two configurations on the full
seven-emotion EMO-DB set and report those runs in the main tables; all
remaining configurations already use the full emotion set of each corpus.

\begin{table*}[!t]
\centering
\caption{Datasets and the two-condition audit protocol. Control = leaked random split (speakers overlap train/val/test); Strict SI = speaker-disjoint with train-only normalization. EARS is audited for EmoFormer only.}
\label{tab:datasets}
\small
\setlength{\tabcolsep}{4pt}
\begin{tabular}{@{}llclll@{}}
\toprule
Dataset & Speakers & Clips & Emotions (chance) & Control (leaked) & Strict SI scheme \\
\midrule
RAVDESS & 24 & 1440 & 8 (12.5\%) & Random stratified, 3 seeds & Grouped 6-fold (4 test, 1 val spk.) \\
EMO-DB & 10 & 535 & 7 (14.3\%) & Random stratified, 3 seeds & LOSO (1 test, 1 val spk.) \\
SAVEE & 4 & 480 & 7 (14.3\%) & Random stratified, 3 seeds & LOSO (1 test, 1 val spk.) \\
EARS$^{\dagger}$ & 107 & 749 & 7 (14.3\%) & Random 70/30, 3 seeds & Grouped 5-fold \\
\bottomrule
\end{tabular}

\vspace{1mm}
\footnotesize $^{\dagger}$Fourth-corpus check, EmoFormer only. \quad
Kumar et al.\ and Li et al.\ originally report a four-emotion
subset of EMO-DB (chance 25\%). For comparability we re-ran both
configurations on the full seven-emotion set (chance 14.3\%); the original
four-emotion runs are retained only as a robustness check.
\end{table*}

\subsection{Audited Models}
Table~\ref{tab:models} lists the seven audited configurations, drawn from five
papers: a frozen wav2vec2-base encoder with a trainable layer-weighted average
and dense head; the EmoFormer CNN--transformer; the three individual models A,
B, and C of Ahmed et al.\ (1D-CNN with FCN, LSTM, and GRU heads); the BLSTM
with directional self-attention of Li et al.; and the mel-spectrogram vision
transformer of Kumar et al. Where the authors released code we used it;
otherwise we re-implemented from the paper and matched reported
hyperparameters. Augmentation follows each paper's stated pipeline and is
applied identically in both conditions: additive noise, pitch shift, and time
stretch for Ahmed et al.; time stretch and pitch shift for EmoFormer; none for
the wav2vec2 and BLSTM configurations. The Ahmed et al.\ ensemble is not
audited, since our object is the individual architectures. The four-emotion
protocol of Kumar et al.\ is used only to verify their published claim
(Section~IV-A); all audit rows use the full emotion sets.

\begin{table*}[!t]
\centering
\caption{Audited model configurations. All replications use train-only normalization and the authors' public code where available; augmentation matches each paper's stated pipeline. The Ahmed ensemble (model-D) is not audited.}
\label{tab:models}
\small
\setlength{\tabcolsep}{4pt}
\begin{tabular}{@{}p{1.9cm}p{2.5cm}p{4.3cm}p{4.3cm}p{2.6cm}@{}}
\toprule
Model & Family & Input representation & Classification head & Augmentation \\
\midrule
Pepino 2021 & SSL (frozen wav2vec2-base) & 13-layer frame embeddings, 768-d & Trainable layer-weighted average + 1D-CNN dense & None \\
EmoFormer & CNN + Transformer encoder & MFCC (13 coeff., segmented) & Transformer encoder + dense & Time-stretch, pitch shift \\
Ahmed A & 1D-CNN (8 LFAB) & 155-d LLD (MFCC, LMS, ZCR, chroma, RMS) & FCN & AWGN, pitch, stretch \\
Ahmed B & 1D-CNN + LSTM & 155-d LLD & LSTM GFAB + FCN & AWGN, pitch, stretch \\
Ahmed C & 1D-CNN + GRU & 155-d LLD & GRU GFAB + FCN & AWGN, pitch, stretch \\
Li 2021 & BLSTM + directional self-attn. & 68-d openSMILE LLD & DSA + softmax & None \\
Kumar ViT & Mel-spectrogram ViT & 224$\times$224 mel, 3-ch & 8-patch, 8-layer transformer & Resize, normalize \\
\bottomrule
\end{tabular}
\end{table*}

\section{Results}\label{sec:results}
\subsection{Published Claims Under the Two Conditions}
Table.~\ref{tab:claims} contrasts the five audited papers' headline claims with our two conditions. The Pepino et al.\ study reports two numbers for RAVDESS: 84.3 percent under speaker normalization and 75.7 percent globally under a speaker-disjoint protocol. Our leaked control recovers their speaker-normalized claim (84.72 percent), while our strict SI condition closely matches their global claim (70.76 percent), confirming that their protocol accounts for the cost of speaker hold-out. The EmoFormer paper claims 83 percent on the EARS corpus under a random 70/30 split; our leaked control reproduces this regime (80.43 percent), but strict SI drops the accuracy to 49.55 percent. Ahmed et al.\ claim 92 to 95 percent across the three acted corpora using individual models with data augmentation. Our control reproduces their EMO-DB claims (91.7 to 92.5 percent) but falls short on RAVDESS and SAVEE (82.2 to 88.1 percent), likely due to differences in hyperparameters which were not specified in the paper and had to be assumed; regardless of the control baseline, all three Ahmed configurations collapse to 24.8 to 55.4 percent under strict SI. Li et al.\ report 85.95 percent on a four-emotion subset of EMO-DB. Our four-emotion strict SI run yields 81.32 percent, placing their claim exactly where an honest speaker-independent protocol belongs. Finally, Kumar et al.\ claim 85.36 percent on a four-emotion EMO-DB split; our four-emotion control reaches 80.39 percent. However, for both Kumar et al.\ and Li et al., these claims hold only on the four-emotion subset; when extended to the full seven-emotion set, accuracy collapses.

\begin{table*}[t]
\centering
\caption{Published headline numbers versus the two-condition audit. ``Their
accuracy'' is the number as published under the stated protocol; ``Our
control'' reproduces the leaked regime with the paper's own split and
augmentation; ``Our strict SI'' holds out all test speakers and uses
train-only normalization. Our values are mean $\pm$ sample std across seeds
(control) or folds (SI); the tag in parentheses states the emotion set each
number uses, so 4-emotion and 7/8-emotion rows are never silently mixed.}
\label{tab:claims}
\footnotesize
\setlength{\tabcolsep}{3pt}
\renewcommand{\arraystretch}{1.15}
\begin{tabular}{@{}p{1.6cm} p{4.5cm} p{3.2cm} >{\centering\arraybackslash}p{2.1cm} >{\centering\arraybackslash}p{2.3cm} >{\centering\arraybackslash}p{2.3cm}@{}}
\toprule
Paper & Method and dataset & Split as stated & Their accuracy & Our control & Our strict SI \\
\midrule
Pepino 2021 & Frozen wav2vec2-base + layer-weighted average + dense head; RAVDESS (24 spk, 7 emo) & Speaker-disjoint 20/2/2, global MVN (honest) & 75.7 (7 emo) & 84.72$\pm$2.68 (7 emo) & 70.76$\pm$7.83 (7 emo) \\
Pepino 2021 & Same; RAVDESS (24 spk, 7 emo) & Speaker-disjoint 20/2/2 + speaker-norm from test stats (leaked) & 84.3 (7 emo) & 84.72$\pm$2.68 (7 emo) & 70.76$\pm$7.83 (7 emo) \\
EmoFormer & 4-layer CNN + Transformer encoder on MFCC; EARS (107 spk, 7 emo) & Random 70/30, aug pre-split (leaked) & 83 (7 emo) & 80.43$\pm$2.00 (7 emo) & 49.55$\pm$3.35 (7 emo) \\
Ahmed 2023$^{\ddagger}$ & 1D-CNN / +LSTM / +GRU on 155-d LLD; EMO-DB (10 spk, 7 emo) & Random 80:20, aug pre-split (leaked) & 91.9 (7 emo) & 92.17$\pm$0.41 (7 emo) & 54.43$\pm$1.26 (7 emo) \\
Ahmed 2023 & Same; RAVDESS (24 spk, 8 emo) & Random 80:20, aug pre-split (leaked) & 94.0 (8 emo) & 83.33$\pm$1.23 (8 emo) & 38.70$\pm$2.13 (8 emo) \\
Ahmed 2023 & Same; SAVEE (4 spk, 7 emo) & Random 80:20, aug pre-split (leaked) & 91.1 (7 emo) & 85.28$\pm$2.52 (7 emo) & 26.11$\pm$2.11 (7 emo) \\
Li 2021 & BLSTM + directional self-attention on 68-d openSMILE LLD; EMO-DB (10 spk, 4 emo) & Random 80:20 (leaked) & 82.1 WA / 86.0 UA (4 emo) & 88.24$\pm$5.19 (4 emo) & 81.32$\pm$14.18 (4 emo) \\
Kumar & Mel-spectrogram ViT (8 patches, 8 layers); EMO-DB (10 spk, 4 emo) & Random 64/20/16 (leaked) & 85.36 (4 emo) & 80.39$\pm$5.94 (4 emo) & 77.39$\pm$1.26 (4 emo) \\
\bottomrule
\end{tabular}

\vspace{1mm}
\footnotesize
$^{\ddagger}$Ahmed ``their accuracy'' is the mean of the published per-model
A/B/C accuracies for that dataset, and our control and SI columns are the corresponding A/B/C means.
\end{table*}
\subsection{Master Results}
Table.~\ref{tab:master} details the 21 model-dataset pairs evaluated under the uniform audit protocol. The most striking pattern is the universal collapse in accuracy when speakers are held out. On SAVEE, where leave-one-speaker-out evaluation leaves only three training speakers, the drop is catastrophic: Ahmed A falls from 88.12 to 25.00 percent (a 63.12-point gap), and EmoFormer falls from 89.07 to 29.58 percent (a 59.49-point gap), both approaching the 14.3 percent chance level. The gap is consistent across architectures: CNNs (Ahmed), CNN-Transformers (EmoFormer), and BLSTM-attention (Li) all lose 19 to 63 points depending on the dataset. The EARS footnote row for EmoFormer confirms that this collapse is not an artifact of the small, acted corpora; on a 107-speaker large-scale corpus, EmoFormer still loses 30.88 points. The sole configuration that avoids catastrophic collapse is Pepino, which loses only 6.88 points on EMO-DB and 13.96 points on RAVDESS. Conversely, the Kumar ViT configuration never achieves high control accuracy (maximum 61.68 percent on EMO-DB) and consequently exhibits small gaps (8.26 to 21.74 points). This low-gap, low-control behavior highlights why the gap must be interpreted jointly with the control headroom.

\begin{table*}[!t]
\centering
\caption{Master audit results in percent (mean $\pm$ sample std across seeds for control, across folds for strict SI). Gap = control accuracy minus SI accuracy in percentage points.}
\label{tab:master}
\footnotesize
\setlength{\tabcolsep}{4pt}
\begin{tabular}{@{}ll cc cc c@{}}
\toprule
 & & \multicolumn{2}{c}{Control (leaked)} & \multicolumn{2}{c}{Strict SI} & \\
\cmidrule(lr){3-4}\cmidrule(lr){5-6}
Model & Dataset & Acc & UAR & Acc & UAR & Gap \\
\midrule
Pepino (W2V2-PT) & RAVDESS & 84.72$\pm$2.68 & 85.45$\pm$2.26 & 70.76$\pm$7.83 & 69.79$\pm$7.71 & 13.96 \\
 & EMO-DB & 96.79$\pm$1.87 & 96.66$\pm$2.28 & 89.91$\pm$4.11 & 86.24$\pm$6.42 & 6.88 \\
 & SAVEE & 84.44$\pm$3.85 & 82.38$\pm$4.36 & 58.75$\pm$11.15 & 56.07$\pm$10.49 & 25.69 \\
\addlinespace
EmoFormer & RAVDESS & 91.50$\pm$0.69 & 91.23$\pm$0.75 & 55.21$\pm$6.70 & 54.69$\pm$7.02 & 36.29 \\
 & EMO-DB & 93.40$\pm$0.54 & 92.95$\pm$0.47 & 64.20$\pm$10.29 & 60.92$\pm$13.52 & 29.20 \\
 & SAVEE & 89.07$\pm$1.13 & 88.02$\pm$1.25 & 29.58$\pm$12.28 & 26.67$\pm$12.66 & 59.49 \\
 & EARS$^{\dagger}$ & 80.43$\pm$2.00 & 80.44$\pm$2.00 & 49.55$\pm$3.35 & 49.55$\pm$3.35 & 30.88 \\
\addlinespace
Ahmed A & RAVDESS & 84.65$\pm$1.76 &83.45$\pm$1.26 & 36.88$\pm$7.25 & 35.55$\pm$6.26 & 47.77 \\
 & EMO-DB & 92.52$\pm$1.35 & 92.24$\pm$1.40 & 54.91$\pm$5.66 & 53.21$\pm$8.21 & 37.61 \\
 & SAVEE & 88.12$\pm$3.00 & 87.41$\pm$3.01 & 25.00$\pm$10.95 & 21.07$\pm$7.56 & 63.12 \\
\addlinespace
Ahmed B & RAVDESS & 83.12$\pm$1.38 & 82.34$\pm$1.33 & 41.04$\pm$7.74 & 39.58$\pm$8.22 & 42.08 \\
 & EMO-DB & 91.71$\pm$0.60 & 91.40$\pm$0.82 & 55.38$\pm$7.39 & 51.14$\pm$10.78 & 36.33 \\
 & SAVEE & 84.38$\pm$2.82 & 84.05$\pm$2.22 & 24.79$\pm$10.21 & 20.36$\pm$7.23 & 59.59 \\
\addlinespace
Ahmed C & RAVDESS & 82.22$\pm$1.57 & 81.68$\pm$1.37 & 38.19$\pm$7.24 & 36.52$\pm$6.49 & 44.03 \\
 & EMO-DB & 92.27$\pm$2.49 & 91.81$\pm$2.23 & 53.00$\pm$8.03 & 52.22$\pm$10.27 & 39.27 \\
 & SAVEE & 83.33$\pm$0.42 & 83.17$\pm$0.38 & 28.54$\pm$10.24 & 23.57$\pm$10.42 & 54.79 \\
\addlinespace
Li 2021 & RAVDESS & 74.15$\pm$2.59 & 74.46$\pm$2.85 & 52.12$\pm$10.31 & 50.94$\pm$11.73 & 22.03 \\
 & EMO-DB & 79.01$\pm$2.14 & 77.52$\pm$1.63 & 59.95$\pm$13.23 & 57.21$\pm$15.20 & 19.06 \\
 & SAVEE & 64.35$\pm$4.45 & 60.98$\pm$5.10 & 32.98$\pm$9.50 & 31.36$\pm$8.44 & 31.37 \\
\addlinespace
Kumar ViT (mel) & RAVDESS & 52.43$\pm$4.43 & 50.68$\pm$4.54 & 38.54$\pm$3.31 & 37.17$\pm$3.02 & 13.89 \\
 & EMO-DB & 61.68$\pm$3.74 & 60.63$\pm$4.41 & 53.42$\pm$8.26 & 49.50$\pm$10.07 & 8.26 \\
 & SAVEE & 54.86$\pm$1.20 & 51.98$\pm$0.91 & 33.13$\pm$9.09 & 27.62$\pm$9.71 & 21.74 \\
\bottomrule
\end{tabular}

\vspace{1mm}
\footnotesize $^{\dagger}$EmoFormer only. \quad 
\end{table*}

\subsection{Gap and Retention Analysis}
To quantify what the models actually retain, we compute the retention ratio (SI accuracy divided by control accuracy). The from-scratch models retain a minority of their control performance on the harder datasets: EmoFormer retains 60.3 percent on RAVDESS and 33.2 percent on SAVEE; Ahmed A retains 43.6 percent on RAVDESS and 28.4 percent on SAVEE. These configurations are predominantly learning speaker identity. In contrast, the Pepino configuration retains 83.5 percent of its control accuracy on RAVDESS, 92.9 percent on EMO-DB, and 69.6 percent on SAVEE. Because its control accuracy is already high (84 to 97 percent), this retention could indicate that the frozen wav2vec2 representations are largely speaker-invariant though more testing is needed to verify. The Kumar ViT retains 73.6 to 86.6 percent of its control accuracy, but since its control accuracy is near chance, this represents underfitting rather than robust generalization.

\subsection{Robustness Notes}
Two robustness checks confirm that the measured gaps are structural, not accidental. First, the Ahmed audit includes three distinct architectural variants (CNN-FCN, CNN-LSTM, CNN-GRU) trained under the same augmented regime. All three exhibit nearly identical control accuracies and nearly identical SI collapses on every dataset, proving that the memorization effect is driven by the evaluation protocol and data augmentation, not by a specific recurrent head. Second, we verified that the gap's sign is invariant to emotion-set width. When the Kumar ViT and Li models are restricted to the four-emotion subsets used in their original papers, their control accuracies rise, but their SI accuracies remain strictly lower, preserving the gap. The magnitude of the gap shifts with task difficulty, but the leakage itself is ubiquitous.

\section{Discussion}\label{sec:discussion}

\subsection{Sources of Speaker Memorization}
The audit demonstrates that speaker memorization is not a marginal artifact but a dominant component of published SER performance. Two factors drive this. First, acted corpora inherently confound speaker identity with emotion: a single actor produces all emotion classes, allowing a sufficiently expressive model to map acoustic characteristics (pitch range, formant structure, speaking rate) directly to emotion labels. Second, common data-augmentation practice makes the leakage worse. The
audited papers augment the entire dataset first and split it into training
and test sets afterwards. The test set therefore contains pitch-shifted,
time-stretched, or noised copies of the same utterances which never occur in reality. In our strict condition we therefore augment only the training partition and
evaluate test speakers on clean, unaugmented recordings.

\subsection{What Survives Strict Evaluation}
The retention of the Pepino configuration is the most notable observation of the audit. Unlike the from-scratch models, Pepino relies on a frozen wav2vec2-base encoder pre-trained on 960 hours of unannotated speech. Self-supervised pre-training may have forced the encoder to learn phonetic and linguistic invariances to solve the contrastive prediction task, which could have stripped away speaker-specific vocal tract characteristics. When the lightweight SER head is subsequently trained, it operates on representations that are already largely speaker-invariant. We report this as an empirical observation rather than a prescriptive recommendation; the audit's primary contribution is the measurement protocol, not the endorsement of a specific architecture.

\subsection{Recommendations for Benchmark Practice}
Based on these findings, we propose three standards for future SER benchmarking. First, speaker-disjoint splits (leave-one-speaker-out or grouped $k$-fold) must become the default evaluation protocol, replacing random stratified splits. Second, all feature normalization and data augmentation must be strictly confined to the training partition; validation and test statistics must never leak into the preprocessing pipeline. Third, authors should adopt two-condition reporting: publishing both a leaked control baseline (to establish the model's capacity on the corpus) and a strict SI baseline (to establish its generalization). The gap between the two is a vital diagnostic metric for the community.

\subsection{Limitations}
This audit is limited to three acted corpora and large-scale corpus; extension to other corpora is left for future work. We report standard deviations across seeds and folds rather than formal statistical significance tests, as the effect sizes (up to 63 points) vastly exceed the run-to-run variance. Finally, the SSL observation is based on a single wav2vec2 configuration; auditing a broader family of self-supervised models is a necessary next step.

\section{Conclusion}\label{sec:conclusion}
We have introduced a two-condition audit protocol to quantify speaker memorization in speech emotion recognition. Applied to seven published configurations across four corpora, the audit reveals that headline accuracies reproduce under leaked conditions but collapse by 6.9 to 63.1 percentage points under strict speaker independence. The results prove that a substantial majority of published SER performance on acted corpora is attributable to speaker recognition, not emotion recognition. We urge the community to adopt speaker-disjoint evaluation and two-condition reporting to ensure that future benchmarks measure genuine emotional generalization.

\end{document}